\documentclass{article}

\usepackage{spconf,amsmath,graphicx,hyperref}
\usepackage[T1]{fontenc}
\usepackage{amsmath,amssymb,amsfonts}
\usepackage{algorithmic}
\usepackage{graphicx}
\usepackage{textcomp}
\usepackage{xcolor}
\usepackage[utf8]{inputenc}
\usepackage{subcaption}  
\usepackage{lipsum} 
\usepackage{float}    
\usepackage{svg}
\usepackage{cuted}
\usepackage[normalem]{ulem}
\usepackage{arydshln}

\title{Plug-and-play forward backward algorithm to restore Landsat images: A preliminary step to uncover the history of surface waters}
\name{Pierre Audisio\textsuperscript{1,2}, Barbara Belletti\textsuperscript{2}, Nelly Pustelnik\textsuperscript{1}}

\address{ \normalsize{
\textsuperscript{1}Ens de Lyon, CNRS, Laboratoire de Physique, F-69342, Lyon, France.}\\
\normalsize{\textsuperscript{2}Université Jean Monnet, CNRS, UMR 5600 Environnement Ville Société, Saint-Etienne, France.}}

\begin{document}
\ninept

\maketitle
\begin{abstract}
The temporal and spatial analysis of river dynamics is a key factor for studying and understanding human impacts on floodplains.
To assess the changes taking place, it is necessary to have high-resolution images with a large spatial coverage and a high temporal revisit frequency over the long term.
Satellite imagery meets several of these criteria. For instance, Sentinel data provide high-resolution images but only after 2015. Therefore, to study water surface evolution prior to this date, it is necessary to rely on other satellite images such as Landsat, which offers longer historical coverage, albeit with lower spatial resolution.  In this study, we aim to increase the spatial resolution of Landsat data from 30 to 10 meters (resolution of Sentinel images). To achieve this goal, we develop an innovative single image super-resolution  method based on a plug-and-play approach.
\end{abstract}

\section{Introduction}
\label{sec:intro}

\noindent \textbf{Context -}  Monitoring river ecosystems is a challenging task due to the complex interactions between ecological and morphological processes. Their complexity lies in their manifestation at multiple spatial scales going from regional or entire watershed to local river sections \cite{habersack2000river}. These ecosystems are particularly vulnerable to long-term changes driven by anthropogenic pressures, such as urbanization, dam construction, and land use changes \cite{Crutzen2002}. 
To effectively capture these dynamics, remote sensing offers a powerful solution. However, accurate monitoring requires satellite data with three key characteristics: large spatial coverage, to capture interdependent regions; high spatial resolution, to detect fine-scale features; and short revisit periods, to track temporal changes.
Satellite multispectral imagery is particularly well suited to this issue, due to the possibility of studying the occupation of land at specific wavelengths, but also thanks to the availability of open-access data at different spatial resolutions, from 10 to 60 meters depending on the satellite. However, high spatial resolution data ($\sim$10 m/pixel) has only become available since the early 2000s.
Consequently, to capture more detailed and extended temporal information, it is necessary to enhance the resolution of satellite images with lower resolutions, such as those from Landsat.

\noindent \textbf{Contribution -} In this study, two sources of data are used. 
Sentinel-2 images, part of the Copernicus project, which started in 2015 \cite{gascon2014}, and Landsat project images that started in 1972 \cite{wulder2019}.
Sentinel-2 images have a spatial resolution of up to 10 meters, while Landsat spectral bands have a resolution of 30 meters.  
The goal of this work is, therefore, to obtain higher resolution images over a longer period of time, and a greater density of images over the years covered simultaneously by Sentinel-2 \cite{gascon2014} and Landsat \cite{claverie2018,wulder2019} satellites.
To do so, we adopt an inverse problem approach to reconstruct high-resolution images from Landsat data ($\sim$30 m/pixel), using Sentinel-2 images ($\sim$10 m/pixel) as the spatial resolution reference. To achieve this goal, we develop a Plug-and-play (PnP) approach, named \texttt{Spect-FB-PnP}, dedicated to multispectral reconstruction, that combines the stability of standard inverse problem formulation and the expressivity of deep neural network models.

\noindent \textbf{Related works -} Single image super-resolution is a well known problem  
that has been solved for many years considering variational approaches including the well-known use of total variation (TV) penalization~\cite{babacan2010variational} or penalization based on wavelet transform \cite{ji2008robust}.
However, despite the effectiveness of these classical techniques, the ability of deep learning to achieve outstanding restoration quality has driven a shift towards neural networks approaches, where model parameters or priors are now learned directly from data  \cite{nazzal2015}.
A wide variety of deep learning frameworks have been made to achieve these tasks \cite{wang2020deep}, starting from supervised model like convolutional neural networks (CNN) based methods \cite{dong2014}, to more recent unsupervised approaches using generative adversarial networks (GAN) 
\cite{ledig2017}. 
Both supervised methods
\cite{wang2021lightweight} and unsupervised method \cite{Prajapati2020}, are able to achieve satisfactory results on super-resolution.
However, on the one hand, supervised approaches requires a significant amount of data for their training, which is particularly challenging with remote sensing data, as supervised methods require manually prepared pairs of high- and low-resolution images.
Additionally, when handling real satellite data, the pairs of high- and low-resolution  images are usually unavailable, because only one satellite image can be acquired at a specific location and time. A common way to handle this problem is to generate low-resolution images using the bicubic interpolation method as a downsampling method \cite{Tuna2018}, but this degradation does not correctly recreate real world low-resolution images \cite{gu2019blind}. 
On the other hand, although unsupervised methods have the significant advantage of not requiring high-resolution data for training, their use in remote sensing imagery remains very limited.
In addition, the more efficient methods using GANs need to make very strong assumptions about the distribution of data for training the prior, which is not always known on remote sensing images, and this can lead to a deterioration in results and the creation of artifacts \cite{daniels2020reducing}. 
Supervised approaches also involve models other than CNNs, such as Transformers \cite{vaswani2017attention}.
The term “Transformer” refers to the self-attention mechanism, which models the global relationships among image regions, eliminating the reliance on content-independent convolution kernels used in CNNs.
One of the most promising methods in image super-resolution using this approach is SwinIR \cite{liang2021swinir}.

Finally, model-based neural network approaches, including unfolded neural networks and Plug-and-Play (PnP) methods, offer a compelling alternative by combining the interpretability of variational methods with the expressiveness of learning-based models, while requiring fewer parameters. PnP approaches rely on splitting algorithms like forward-backward or ADMM  \cite{Venkatakrishnan2013,park2025plug,hurault2024convergent} to separate the action of the data fidelity and prior terms. This separation enables the use of learned CNNs as priors without needing pair of low- and high-resolution. 
In remote sensing, super-resolution techniques meet a great interest \cite{wang2022}, however, model-based neural networks like  PnP have been poorly explored. To the best of our knowledge, the only preliminary study was conducted using the U.S. Geological Survey National Map Urban Area imagery collection data \cite{tao2020}, focusing on RGB images. In the present contribution, we extend this framework to multispectral image enhancement and deeply explore the performance on real data. 

\noindent\textbf{Outline -} Section~\ref{sec:method} describes the specificities of the considered satellite images, the forward model that establishes the relation between Landsat and Sentinel images, and the adopted PnP strategy with a specific focus on the denoiser.
In Section~\ref{sec:dataset}, the description of the two considered datasets (synthetic and real) is provided as well as the calibration we made for the forward model. In Section~\ref{sec:numerical_experiment}, we present a first set of experiments, establishing the behaviour of the proposed PnP w.r.t hyperparameters such as the step-size involved in FB-PnP but also the regularization parameter. We then report the reconstruction results obtained on the two different datasets, along with the comparison of \texttt{Spec-FB-PnP} with both a classical interpolation method and the advanced learning-based approach SwinIR.
Finally, in Section~\ref{sec:conclusion}, we conclude by highlighting the advantages of \texttt{Spec-FB-PnP} and its futur use to uncover the history of surface water.
\vspace{-0.2cm}

\section{Proposed method: \texttt{Spec-FB-PnP}}
\label{sec:method}
\vspace{-0.2cm}

\noindent \textbf{Data -} In this study two sources of data are used. 
Sentinel-2 images, part of the Copernicus project which started in 2015 \cite{gascon2014}, and Landsat project which began in 1972 \cite{wulder2019}.
Sentinel-2 images have a spatial resolution of up to 10m, while Landsat-8/9 images have a resolution of 30m. Both satellites cover a wide range of spectral bands (11 for Landsat-8/9 and 12 for Sentinel-2), however in this study we only focus on green, blue, red, and near-infrared bands for their ability to highlight elements of interest in land use and the proximity of their central wavelengths on both data sources. If the reconstruction quality will be of importance in this study, another quantity will also be considered: the normalized difference water index (NDWI), which highlights water in the images. This spectral index allows extracting hydromorphological metrics, such as the water channel perimeter or the surface area of open water. This index, is calculated from the “green” and “NIR” spectral bands. In the case of Sentinel-2 data (denoted $\mathrm{x}\in \mathbb{R}^{BN}$ where $B$ denotes the number of bands and $N$ the number of pixels), it can be written as follows, for every pixel $n$,  
\begin{equation}
\label{eq:ndwi}
\mathrm{d}_n^{\mathrm{NDWI}} = \frac{\mathrm{x}_{\mathrm{green},n} -\mathrm{x}_{\mathrm{NIR},n}}{\mathrm{x}_{\mathrm{green},n} + \mathrm{x}_{\mathrm{NIR},n}}.
\end{equation}

\noindent \textbf{Forward model -} Following \cite{wang2020deep}, we formulate the forward model as, for every band $b$,\vspace{-0.2cm}
\begin{equation}
	\mathrm{z}_b = (\phi*\bar{\mathrm{x}}_b)\downarrow_s + \varepsilon
	\label{eq:forward}
\end{equation}
where $\bar{\mathrm{x}} = (\bar{\mathrm{x}}_b)_{1\leq b\leq B}\in\mathbb{R}^{BN}$ denotes the original high resolution (HR) image (i.e., Sentinel-2 image), $\phi$ denotes the blurring kernel, $\downarrow_s$ the downsampling operator with a decimation factor $s =3$, $\varepsilon$ the stochastic degradation associated with Gaussian noise, and $\mathrm{z}\in \mathbb{R}^{BM}$ the low-resolution (LR) observation (i.e., Landsat-8/9). 

Our goal is to recover a HR estimate $\widehat{\mathrm{x}}$ close from $\bar{\mathrm{x}}$ considering the LR image $\mathrm{z}$. The long-term goal being to densify the HR observations on the time period where both Landsat and Sentinel-2 images are available (i.e., 2015-now) but also obtain HR images on the period 1972-2015 where only LR Landsat-8/9 images are available. 

In the following, the forward model will be simply formulated as 
$\mathrm{z} = \mathrm{A} \bar{\mathrm{x}} + \varepsilon$ 
where $\mathrm{A}$ encodes both the convolution with $\phi$ and the downsampling.

\noindent \textbf{Plug-and-play -} The literature of model-based neural network including unfolded neural network and Plug-and-play methods have significantly increased in the past ten years and is now established as a good compromise between standard neural networks and variational approaches. Specifically, we investigate the well-established Forward-Backward PnP (FB-PnP) framework \cite{hurault2022proximal,pesquet2021learning}. While FB-PnP has been widely studied for RGB image processing tasks, its use in multispectral imagery remains limited, mainly because it requires the development of a dedicated denoiser capable of handling multispectral data.

The iterations of FB-PnP read:
\begin{equation}
\label{eq:spec-fb-pnp}
{\mathrm{x}}^{[k+1]} = \mathbf{D}_{\theta,\tau\lambda} (\mathrm{x}^{[k]} - \tau \mathrm{A}^\top (\mathrm{A} \mathrm{x}^{[k]} - \mathrm{z})) 
\end{equation}
where $\tau>0$ is the step-size, $\lambda$ a regularization parameter, $\theta$ the neural networks parameters, and $\mathbf{D}_{\theta,\tau\lambda}$ the denoiser. When $\mathbf{D}_{\theta,\tau\lambda} = \mathrm{prox}_{\tau \lambda g}$ we recover standard FB schemes and convergence to $\widehat{\mathrm{x}} \in \mathrm{Argmin}_ \mathrm{x} \frac{1}{2} \Vert \mathrm{A} \mathrm{x} - \mathrm{z} \Vert_2^2 + \lambda g(\mathrm{x})$ is guaranteed for specific choice of $\tau$ \cite{combettes2005signal}. Under technical assumptions, it has been established that the sequence $({\mathrm{x}}^{[k]})_{k=0,1,\ldots}$ converge to $\widehat{\mathrm{x}}$ such that $0\in \mathrm{A}^\top (\mathrm{A} \widehat{\mathrm{x}} - \mathrm{z}) + \mathbf{T}(\widehat{\mathrm{x}})$, where $\mathbf{T}$ is related to $\mathbf{D}_{\theta,\tau\lambda}$. The involved denoiser should insure specific properties that can be imposed during the training, often at a price of reduced performance \cite{hurault2022proximal,pesquet2021learning}.

\noindent \textbf{Building the denoiser -} For this work, we choose for the denoiser $\mathbf{D}_{\theta,\tau\lambda}$ a DRUNet \cite{zhang2021}, which is a combination of the convolutional network U-Net \cite{ronneberger2015u}, and the deep residual learning network ResNet \cite{he2016deep}. Such a denoiser is available in DeepInverse library \cite{tachella2025deepinverse} but for RGB image denoising. To design $\mathbf{D}_{\theta,\tau\lambda}$ for multispectral image denoising, it was first necessary to change the first and last layers to the source code in order to enable the network to train on 4-channel multispectral data and configure its learning accordingly.

\begin{figure}[htbp]
	\centering
	\begin{subfigure}[b]{0.09\textwidth}
		\includegraphics[width=\textwidth]{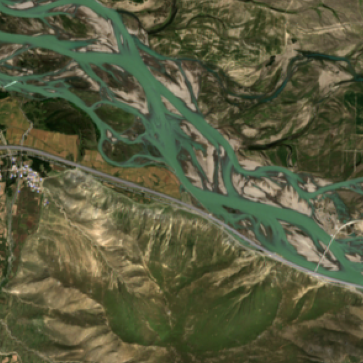}
	\end{subfigure}
	\begin{subfigure}[b]{0.09\textwidth}
		\includegraphics[width=\textwidth]{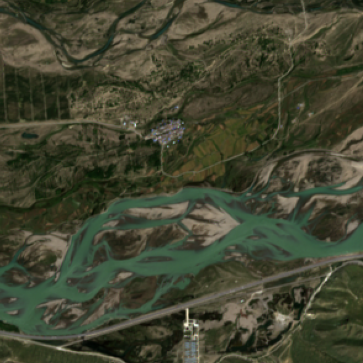}
	\end{subfigure}
	\begin{subfigure}[b]{0.09\textwidth}
		\includegraphics[width=\textwidth]{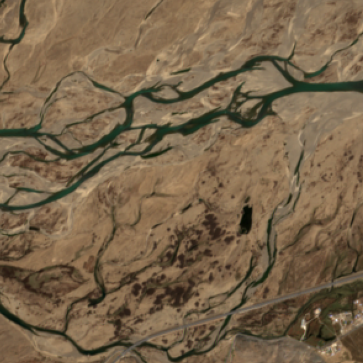}
	\end{subfigure}
	\begin{subfigure}[b]{0.09\textwidth}
		\includegraphics[width=\textwidth]{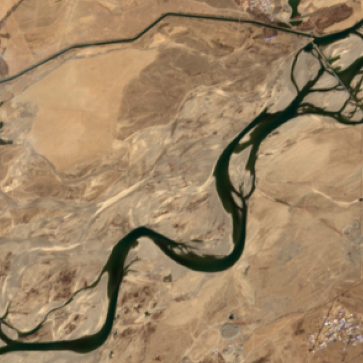}
	\end{subfigure}
	
	
	\begin{subfigure}[b]{0.09\textwidth}
		\includegraphics[width=\textwidth]{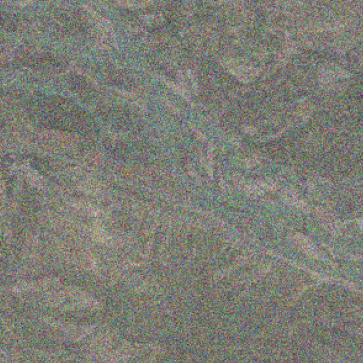}
	\end{subfigure}
	\begin{subfigure}[b]{0.09\textwidth}
		\includegraphics[width=\textwidth]{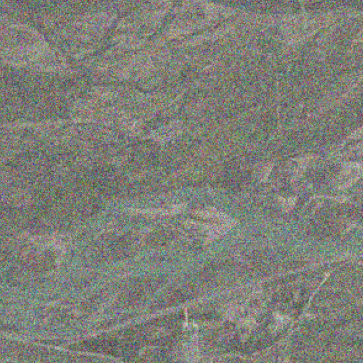}
	\end{subfigure}
	\begin{subfigure}[b]{0.09\textwidth}
		\includegraphics[width=\textwidth]{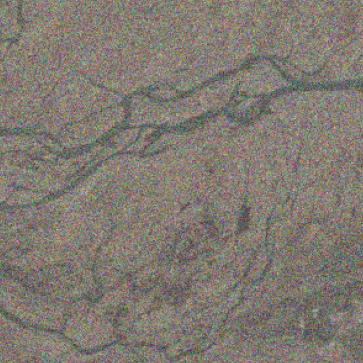}
	\end{subfigure}
	\begin{subfigure}[b]{0.09\textwidth}
		\includegraphics[width=\textwidth]{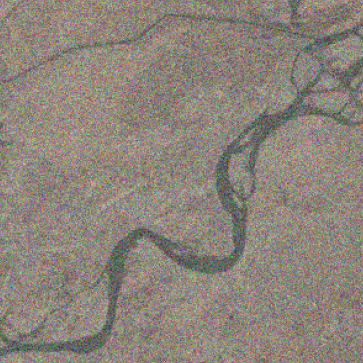}
	\end{subfigure}
	
	\caption{DRUNet training dataset: Original images $\overline{\mathrm{x}}^{(\ell)}$ and noisy versions $\mathrm{y}^{(\ell)}$ with a Gaussian noise with $\sigma=0.09$.}
	\label{fig1}\vspace{-0.2cm}
\end{figure}

To learn the hyperparameters $\theta$ of the proposed denoiser $\mathbf{D}_{\theta,\tau\lambda}$, we created a dataset $\mathcal{D} = \{(\overline{\mathrm{x}}^{(\ell)},\mathrm{y}^{(\ell)})_{\ell=\{1,\ldots,L_D\}}\}$ such that $\mathrm{y}^{(\ell)} = \overline{\mathrm{x}}^{(\ell)} + \epsilon$, where $\overline{\mathrm{x}}^{(\ell)}$ denotes the HR Sentinel-2 images and $\epsilon$ a white Gaussian noise with standard deviation $\sigma$. These $\overline{\mathrm{x}}^{(\ell)}$ images were obtained using $N = 300 \times 300$ pixels crops of Sentinel-2 images centered on the river corridor. The studied site selected for this work is the Lhasa River in Tibet. This area is of particular interest as it has experienced rapid urbanization in recent years, significantly impacting the development of the alluvial plain and continuing to shape it over time. Sentinel-2 images are acquired over the same area of the Earth's surface every 5 days, since data began to be accessible online in 2016. The resulting dataset is composed of  $L_D = 1200$ images. Few examples from $\mathcal{D}$ are displayed in Figure~\ref{fig1}. \vspace{-0.2cm}

We consider AdamW optimizer with a learning rate of 0.001. A StepLR scheduler with a step size of 100 and a decay factor of 0.5. The network was trained for 1500 epochs.
The learning rate and decay factor were chosen based on tests conducted in the Deep Inverse project \cite{tachella2025deepinverse} using three-channel images. To evaluate the impact of the batch size used during the training, two different batch-size have been employed: 4 and 17. In this article, we will only present the reconstruction obtained with a batch-size of 4, given the best results obtained.

\texttt{Spec-FB-PnP} refers to \eqref{eq:spec-fb-pnp} when $\mathbf{D}_{\theta,\tau\lambda}$ is the denoiser described above.

\vspace{-0.3cm}
\section{Datasets and forward model calibration}
\label{sec:dataset}
\vspace{-0.3cm}
 
To evaluate the performance of the \texttt{Spec-FB-PnP} algorithm for increasing the resolution of Landsat-8/9 images to Sentinel-2 images, we consider two types of datasets: a synthetic dataset and a real one. On the one hand, the synthetic dataset is created to mimic as close as possible  the real one and to evaluate the performance of \texttt{Spec-FB-PnP} on a controlled dataset, particularly to verify whether any artifacts are introduced. On the other hand, evaluating \texttt{Spec-FB-PnP} on the real dataset allows us to check its stability on a non-ideal situation. \\
\noindent\textbf{Real dataset and calibration of the forward model -} This dataset consists of pairs of Sentinel-2 ($\overline{\mathrm{x}}_\ell$) and Landsat ($\mathrm{z}_\ell^{\textrm{real}}$) images, such that $\mathcal{R}=\{(\overline{\mathrm{x}}_\ell,\mathrm{z}_\ell^{\textrm{real}})_{\ell=\{1,\ldots,L_R\}}\}$. The underlying forward model is of the form \eqref{eq:forward} where $\phi$, $s$, and $\varepsilon$ are estimated from the data. It is also important to notice that because these data come from different sources, it was not possible to obtain acquisitions that were spatially aligned on the same location and at the same time.
To overcome these constraints, after downloading the images from the two sources, we classified them chronologically in order to create $\{(\overline{\mathrm{x}}_\ell,\mathrm{z}_\ell^{\textrm{real}})\}$ image pairs that were as close as possible in their acquisition time (maximum gap of three days).
For spatial alignment, it was possible to use Landsat data from the Harmonized Landsat Sentinel-2 (HLS) project \cite{claverie2018}, which reprojected Landsat images onto the Sentinel-2 UTM tiling grid.

The $\overline{\mathrm{x}}_\ell$ images were obtained by cropping Sentinel-2 images of size $11000 \times 11000$ pixels using a $N = 300 \times 300$ pixels window, focusing the extraction on the river corridor using the associated centerline obtained from the Global Surface Water (GSW) dataset \cite{pekel2016high}.
The same procedure was used on Landsat images of size $3660 \times 3660$ pixels using a $M = 100 \times 100$ pixels window. 
This difference in window size is due to the scale factor $s=3$. The blurring kernel $\phi$ with a standard deviation $\sigma=0.09$ and a size of $4\times4$ pixels,   has been calibrated by taking into account concepts from the literature for the choice of kernel width \cite{efrat2013accurate} and from $(\overline{\mathrm{x}}_\ell,\mathrm{z}_\ell^{\textrm{real}})$ in order minimize the error $\Vert (\phi *\overline{\mathrm{x}}_\ell - \mathrm{z}_\ell^{\textrm{real}})\downarrow_s \Vert_2^2$.  The standard deviation of white Gaussian noise, was estimated using the median absolute deviation technique applied on wavelet details coefficients of the Landsat image: $\sigma\approx\frac{\text{median}\left(\left|\mathrm{W}\mathrm{z}_\ell^{\textrm{real}} \right| \right)}{0.6745}$ \cite{mallat1999wavelet}.

\noindent\textbf{Synthetic dataset -}  This second dataset has been defined to validate the \texttt{Spec-FB-PnP} on data for which ground truth is available. The dataset is $\mathcal{S}=\{(\overline{\mathrm{x}}_\ell,\mathrm{z}_\ell^{\textrm{synt}})_{\ell=\{1,\ldots,L_S\}}\}$ with $\mathrm{z}_\ell^{\textrm{synt}} = (\phi * \overline{\mathrm{x}}_\ell)\downarrow_s + \varepsilon$ where $\phi$ and $\varepsilon$ are calibrated from the real Landsat images $\mathrm{z}_\ell^{\textrm{real}}$ as described in the paragraph above. 

\begin{figure}[h]
	\centering
    \begin{tabular}{p{1.7cm}p{1.7cm}p{1.7cm}p{1.7cm}}
    \includegraphics[height=1.9cm]{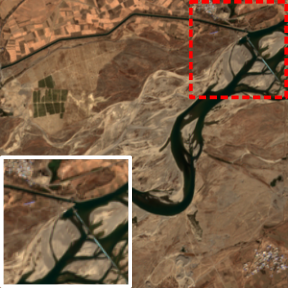}
	& \includegraphics[height=1.9cm]{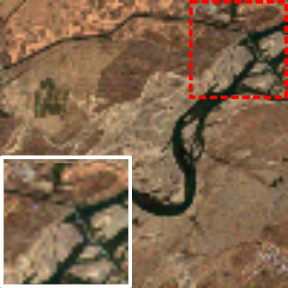}
	&\includegraphics[height=1.9cm]{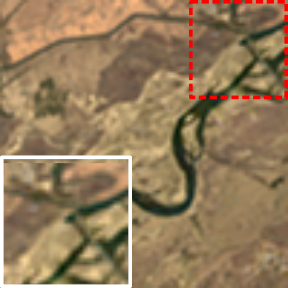}
	\\
	\multicolumn{1}{c}{\footnotesize{HR Sentinel-2 $\overline{\mathrm{x}}_\ell$}} 
	&\multicolumn{1}{c}{\footnotesize{LR Synthetic $\mathrm{z}_\ell^{\textrm{synt}}$}} 
	&\multicolumn{1}{c}{\footnotesize{LR Landsat-8/9 $\mathrm{z}_\ell^{\textrm{real}}$}}\\	

    \includegraphics[height=1.9cm]{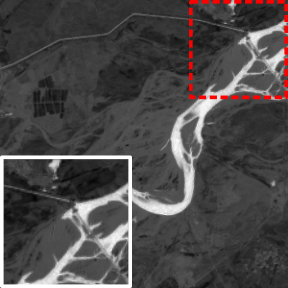}
	& \includegraphics[height=1.9cm]{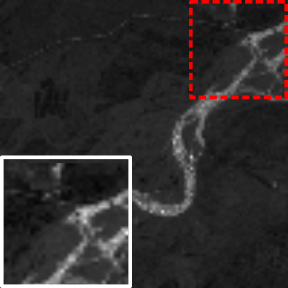}
	&\includegraphics[height=1.9cm]{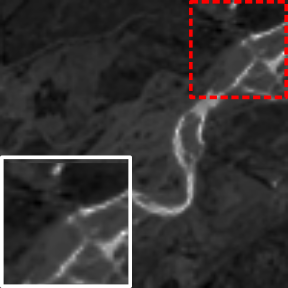}
	\\
	\multicolumn{1}{c}{\footnotesize{NDWI}} 
	&\multicolumn{1}{c}{\footnotesize{NDWI}} 
	&\multicolumn{1}{c}{\footnotesize{NDWI}}\\
    \includegraphics[height=1.9cm]{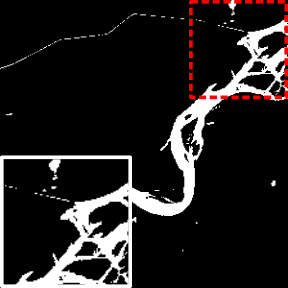}
	& \includegraphics[height=1.9cm]{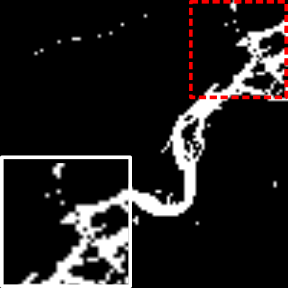}
	&\includegraphics[height=1.9cm]{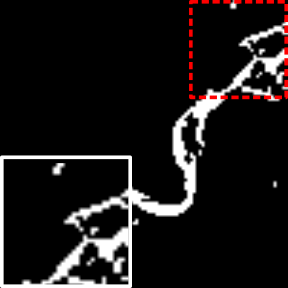}
	\\
	\multicolumn{1}{c}{\footnotesize{Threshold NDWI}} 
	&\multicolumn{1}{c}{\footnotesize{Threshold NDWI}} 
	&\multicolumn{1}{c}{\footnotesize{Threshold NDWI}}\\
	\end{tabular}
	\caption{Row 1: Sentinel-2, synthetic Landsat, and real Landsat images. Row 2: associated NDWI index images. Row 3: thresholded NDWI index allowing water surface extraction.}
	\label{fig2}
\end{figure}

An illustration of images extracted from these two datasets and associated NDWI images is provided in Figure~\ref{fig2}.

\section{Numerical experiment}
\label{sec:numerical_experiment}
\noindent \textbf{Evaluation criteria -} 
The PSNR and SSIM are used as indicators of reconstruction quality. We also compute the surface area of open water by thresholding the NDWI index \eqref{eq:ndwi} (cf. Figure~\ref{fig2}, third line).

\begin{figure}[h]
\centering
    \includegraphics[height= 2.8cm,width=8.5cm]{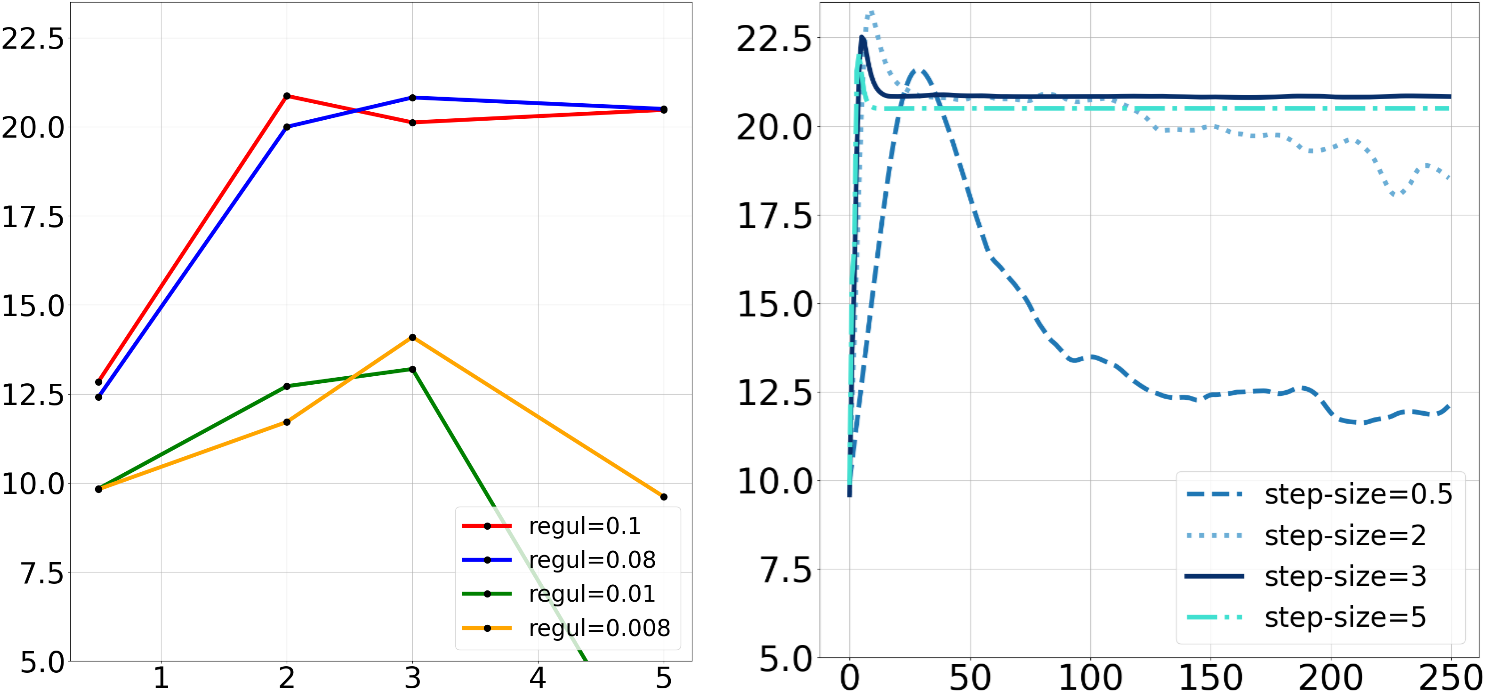} 
	\caption{Evaluation of the choice of regularization parameter $\lambda$ and step size $\tau$ in \texttt{Spec-FB-PnP}. Left: PSNR as a function of step size with $\tau$ a  fixed  regularization term $\lambda$. Right: Evolution  of  PSNR w.r.t iterations for $\lambda=0.008$  and  different step-sizes. }\vspace{-0.3cm}
	\label{fig:param}
\end{figure}

\noindent \textbf{Comparisons with state-of-the-art methods -}  The reconstructions obtained with the Plug-and-Play method are compared with those from the bicubic interpolation implemented in PyTorch, as well as the neural network-based method named SwinIR \cite{liang2021swinir}.  
This second methodology was trained on the same dataset as the denoising DRUNet, except that the network takes also into account the change of resolution from Landsat-8/9 to Sentinel-2. The number of epochs for the training is the same than for our denoiser (i.e. 1500).  \\
\textbf{Impact of the step-size and regularization parameter in PnP -} In Figure~\ref{fig:param}(left), we display the PSNR obtained with the proposed PnP method for different choices of step-size $\tau$ and regularization parameter $\lambda$. For the optimal choice of regularization parameter (i.e. 0.08) we display in Figure~\ref{fig:param}(right) the evolution of the PSNR w.r.t iterations in order to observe the convergence of the algorithm for specific step-size choices. We clearly see the configuration with the step-size $\tau=3$ leads to the best reconstruction results and leads to a converging sequence in terms of PSNR. In the following experiments the choice of $\tau=3$ and $\lambda=0.08$ is made.

\begin{table}[h]
\centering
\small
\begin{tabular}{|l|c|c|c|}
\hline
\textbf{Method} & \textbf{SSIM} & \textbf{PSNR (dB)} & \textbf{Water Area (m$^2$)} \\
\hline

\multicolumn{4}{|c|}{\textbf{Image A – Sentinel Reference (22\,580 m$^2$)}} \\
\hline
Synthetic A              &  –    &  –    & 8\,730 \\
Bicubic interpolation     & 0.59  & 22.7  & 21\,340 \\
SwinIR                    & 0.87  & \textbf{30.0} & 21\,980 \\
PnP                       & \textbf{0.88} & 28.4 & \textbf{22\,620} \\
\hline

\multicolumn{4}{|c|}{\textbf{Image A – Landsat Reference (4\,740 m$^2$)}} \\
\hline
Bicubic interpolation     & 0.51  & 19.1  & 15\,730 \\
SwinIR                    & \textbf{0.65} & \textbf{20.3} & \textbf{16\,600} \\
PnP                       & 0.60  & 19.5  & 15\,510 \\
\hline

\multicolumn{4}{|c|}{\textbf{Image B – Sentinel Reference (26\,620 m$^2$)}} \\
\hline
Synthetic B              &  –    &  –    & 7\,290 \\
Bicubic interpolation     & 0.62  & 24.2  & \textbf{26\,310} \\
SwinIR                    & \textbf{0.85} & \textbf{30.8} & 24\,880 \\
PnP                       & 0.82  & 27.6  & 23\,670 \\
\hline

\multicolumn{4}{|c|}{\textbf{Image B – Landsat Reference (7\,110 m$^2$)}} \\
\hline
Bicubic interpolation     & 0.53  & 14.73 & \textbf{21\,920} \\
SwinIR                    & 0.60  & 14.94 & 19\,010 \\
PnP                       & \textbf{0.64} & \textbf{15.03} & 20\,530 \\
\hline

\end{tabular}
\caption{Quantitative evaluation of reconstruction performance for Images A and B on synthetic and real data. Best values per section are shown in bold.}
\label{tabular_image_AB}
\end{table}

\begin{figure}[t]
	
	\begin{tabular}{p{1.7cm}p{1.7cm}p{1.7cm}p{1.7cm}}
		\textbf{Sentinel A}
		&\textbf{(Zoom)}
		&\textbf{Sentinel B}
		&\textbf{(Zoom)}\\
		\includegraphics[height=1.9cm]{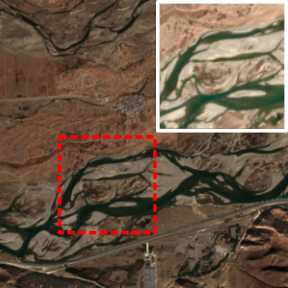}
		&\includegraphics[height=1.9cm]{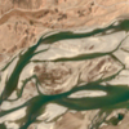}
		& \includegraphics[height=1.9cm]{images/figure2/ref_RGB.png}
		& \includegraphics[height=1.9cm]{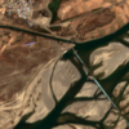}\\
		\hline	 
		\multicolumn{4}{c}{\textbf{Observation $\mathrm{z}$} }\\
		 \multicolumn{1}{c}{\textbf{Synthetic}}
		&\multicolumn{1}{c}{\textbf{Landsat}} 
		&\multicolumn{1}{c}{Synthetic B} 
		&\multicolumn{1}{c}{Landsat B}\\		
		\includegraphics[height=1.9cm]{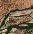}
		&\includegraphics[height=1.9cm]{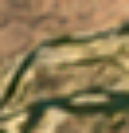}
		& \includegraphics[height=1.9cm]{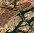}    
		&\includegraphics[height=1.9cm]{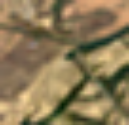}\\
		\multicolumn{4}{c}{\textbf{Reconstruction Bicubic} }\\	
		\includegraphics[height=1.9cm]{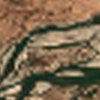}
		&\includegraphics[height=1.9cm]{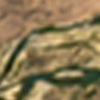}
		&\includegraphics[height=1.9cm]{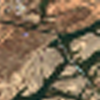}
		&\includegraphics[height=1.9cm]{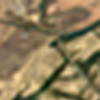}\\
		\multicolumn{4}{c}{\textbf{Reconstruction Swinir} }\\
		\includegraphics[height=1.9cm]{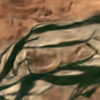}	
		&\includegraphics[height=1.9cm]{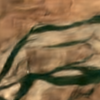}
		&\includegraphics[height=1.9cm]{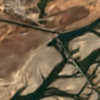}
		&\includegraphics[height=1.9cm]{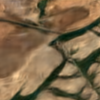}\\
		\multicolumn{4}{c}{\textbf{Reconstruction PnP} }\\	
		\includegraphics[height=1.9cm]{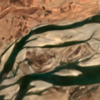}
		&\includegraphics[height=1.9cm]{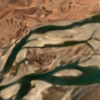}
		&\includegraphics[height=1.9cm]{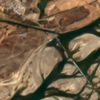}
		&\includegraphics[height=1.9cm]{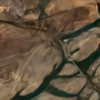}\\
		
	\end{tabular} 
	\caption{Reconstruction of two sections of the Lhasa River (Tibet) using bicubic, SwinIR, and PnP methods, based on both real Landsat images and synthetic Landsat images (degraded Sentinel-2).}\vspace{-0.5cm}
	\label{fig:reconstruction}
\end{figure}

\noindent \textbf{Comments on reconstruction performance -} 
Figure~\ref{fig:reconstruction} presents two sections of the Lhasa River (referred as A and B) as well as a cropped version on each, shown in the first row. These images are particularly well suited for evaluating the reconstruction results because, although the maximum temporal gap between the acquisitions is about three days, no significant changes occurred in that interval on the real data. 

On the second row, we display the associated synthetic images, obtained by degrading Sentinel-2 data to match the spatial resolution and quality of Landsat-8/9 images (denoted Synthetic A/B), and the corresponding real Landsat-8/9 images. The reconstruction results are then displayed on row 3 to 5. Table~\ref{tabular_image_AB} presents the indicators extracted from the reconstructions.

A first visual assessment of the reconstructions reveals that the results obtained with methods based on neural networks (SwinIR and PnP) outperform the classic bicubic interpolation method.
This is true for reconstructions based on synthetic images as well as Landsat images.
As expected, it can also be noted that the results obtained via the reconstruction of synthetic images exceed the scores obtained from real Landsat images in all metrics (see Tabular~\ref{tabular_image_AB}). 
Looking more closely at the reconstructions obtained with SwinIR and PnP, we can see that most of the river branches have been correctly reconstructed, despite visible smoothing in areas outside the watercourse, which have lost detail, particularly in the reconstruction of Landsat A and B by SwinIR.
In terms of water area estimation, we can see that all methods allows to significantly improve the estimation of water area, highlighting the benefit of performing reconstruction of Landsat-8/9 images.

The objective is not only to obtain the most accurate overall measurement of water surface area, but rather to achieve a good compromise between image reconstruction of the spatial structures where water is present.
Such structural information is crucial, as it enables the assessment of how the watercourse evolves over time.
The SSIM score measures the structural similarity between a reference image and its reconstruction, and thus accurately reflects the point we have just made.
Table~\ref{tabular_image_AB} shows that the PnP method achieves the highest SSIM score for the reconstructions of Synthetic A and Landsat B. In the other two cases (Synthetic B and Landsat A), SwinIR obtains the best results, with a maximum difference of only 0.05 points with \texttt{Spec-FB-PnP}, which is negligible.

It is also important to notice that \texttt{Spec-FB-PnP} is twice faster than SwinIR.

\vspace{-0.3cm}
\section{Conclusion}
\label{sec:conclusion}
\vspace{-0.2cm}

In this study, we propose a new \texttt{Spec-FB-PnP} method to retrieve the information lost on Landsat 8/9  compared to Sentinel-2 images. The method is evaluated against a neural network solution SwinIR and a the  classic bicubic interpolation.
The performance has been evaluated in terms of reconstruction quality and hydromorphologically quantities on the Lhasa River in Tibet.
The SSIM score shows that, on average, the \texttt{Spec-FB-PnP} method offers a good compromise to achieve a reconstruction of the structures  with a good estimation of water area estimation.
These results open up prospects for the use of the \texttt{Spec-FB-PnP} method more systematically to incrase the resolution of Landsat image from 1972 to the present day and on other watercourses around the world where training data is very limited due to cloud cover (e.g., the Amazon River).

\vspace{-0.2cm}

\section{Acknowledgements}
\vspace{-0.2cm}

\noindent This work is funded as part of ANR GloUrb project (ANR-22-CE03-0005). 
The project is also co-financed by the Labex IMU (ANR-10-LABEX-0088) and the EUR H2O'Lyon (ANR-17-EURE-0018) of the University of Lyon, as part of the "Investissements d'Avenir" programme operated by the ANR.

\vfill\pagebreak

\bibliographystyle{IEEEbib}
\bibliography{Pierre_Audisio_ICASSP2026}

\begin{thebibliography}{10}

\bibitem{habersack2000river}
H.~M. Habersack,
\newblock ``The river-scaling concept ({RSC}): a basis for ecological
  assessments,''
\newblock {\em Hydrobiologia}, vol. 422, pp. 49--60, 2000.

\bibitem{Crutzen2002}
P.~J. Crutzen,
\newblock ``The {A}nthropocene,''
\newblock {\em J. de Physique IV}, 2002.

\bibitem{gascon2014}
F.~Gascon, E.~Cadau, O.~Colin, B.~Hoersch, C.~Isola, B.~L. Fernandez, and
  P.~Martimort,
\newblock ``Copernicus {S}entinel-2 mission: products, algorithms and
  cal/val,''
\newblock in {\em Earth observing systems XIX}. SPIE, 2014, vol. 9218, pp.
  455--463.

\bibitem{wulder2019}
M.~A. Wulder, T.~R. Loveland, D.~P. Roy, C.~J. Crawford, J.~G. Masek, C.~E.
  Woodcock, R.~G. Allen, M.~C. Anderson, A.~S. Belward, and W.~B. Cohen,
\newblock ``Current status of landsat program, science, and applications,''
\newblock {\em Remote Sensing of Environment}, vol. 225, pp. 127--147, 2019.

\bibitem{claverie2018}
M.~Claverie, J.~Ju, J.~G. Masek, J.~L. Dungan, E.~F. Vermote, J.-C. Roger,
  S.~V. Skakun, and C.~Justice,
\newblock ``The harmonized {L}andsat and {S}entinel-2 surface reflectance data
  set,''
\newblock {\em Remote Sensing of Environment}, vol. 219, pp. 145--161, 2018.

\bibitem{babacan2010variational}
S.~D. Babacan, R.~Molina, and A.~K. Katsaggelos,
\newblock ``Variational bayesian super resolution,''
\newblock {\em IEEE Trans. Image Process.}, vol. 20, no. 4, pp. 984--999, 2010.

\bibitem{ji2008robust}
H.~Ji and C.~Ferm{\"u}ller,
\newblock ``Robust wavelet-based super-resolution reconstruction: theory and
  algorithm,''
\newblock {\em IEEE Trans. Pattern Anal. Match. Int.}, vol. 31, no. 4, pp.
  649--660, 2008.

\bibitem{nazzal2015}
M.~Nazzal and H.~Ozkaramanli,
\newblock ``Wavelet domain dictionary learning-based single image
  superresolution,''
\newblock {\em Signal, Image and Video Processing}, vol. 9, no. 7, pp.
  1491--1501, 2015.

\bibitem{wang2020deep}
Z.~Wang, J.~Chen, and S.~C.~H. Hoi,
\newblock ``Deep learning for image super-resolution: A survey,''
\newblock {\em IEEE Trans. Pattern Anal. Match. Int.}, vol. 43, no. 10, pp.
  3365--3387, 2020.

\bibitem{dong2014}
C.~Dong, C.~C. Loy, K.~He, and X.~Tang,
\newblock ``Learning a deep convolutional network for image super-resolution,''
\newblock in {\em European Conference on Computer Vision}. Springer, 2014, pp.
  184--199.

\bibitem{ledig2017}
C.~Ledig, L.~Theis, F.~Huszár, J.~Caballero, A.~Cunningham, A.~Acosta,
  A.~Aitken, A.~Tejani, J.~Totz, and Z.~Wang,
\newblock ``Photo-realistic single image super-resolution using a generative
  adversarial network,''
\newblock in {\em IEEE Conf. Comput. Vis. Pattern Recognit.}, 2017, pp.
  4681--4690.

\bibitem{wang2021lightweight}
J.~Wang, Y.~Wu, L.~Wang, L.~Wang, O.~Alfarraj, and A.~Tolba,
\newblock ``Lightweight feedback convolution neural network for remote sensing
  images super-resolution,''
\newblock {\em IEEE Access}, vol. 9, pp. 15992--16003, 2021.

\bibitem{Prajapati2020}
K.~Prajapati, V.~Chudasama, H.~Patel, K.~Upla, R.~Ramachandra, K.~Raja, and
  C.~Busch,
\newblock ``Unsupervised single image super-resolution network ({USISResNet})
  for real-world data using generative adversarial network,''
\newblock in {\em IEEE/CVF Conf. Comput. Vis. Pattern Recognit.}, 2020, pp.
  464--465.

\bibitem{Tuna2018}
C.~Tuna, G.~Unal, and E.~Sertel,
\newblock ``Single-frame super resolution of remote-sensing images by
  convolutional neural networks,''
\newblock {\em International Journal of Remote Sensing}, vol. 39, no. 8, pp.
  2463--2479, 2018.

\bibitem{gu2019blind}
J.~Gu, H.~Lu, W.~Zuo, and C.~Dong,
\newblock ``Blind super-resolution with iterative kernel correction,''
\newblock in {\em IEEE/CVF Conf. Comput. Vis. Pattern Recognit.}, 2019, pp.
  1604--1613.

\bibitem{daniels2020reducing}
M.~Daniels, P.~Hand, and R.~Heckel,
\newblock ``Reducing the representation error of gan image priors using the
  deep decoder,''
\newblock {\em arXiv preprint arXiv:2001.08747}, 2020.

\bibitem{vaswani2017attention}
A.~Vaswani, N.~Shazeer, N.~Parmar, J.~Uszkoreit, L.~Jones, A.~N. Gomez,
  {\L}.~Kaiser, and I.~Polosukhin,
\newblock ``Attention is all you need,''
\newblock {\em Adv. Neural Inf. Process. Syst.}, vol. 30, 2017.

\bibitem{liang2021swinir}
J.~Liang, J.~Cao, G.~Sun, K.~Zhang, L.~Van~Gool, and R.~Timofte,
\newblock ``Swin{IR}: Image restoration using swin transformer,''
\newblock in {\em IEEE/CVF International Conference on Computer Vision}, 2021,
  pp. 1833--1844.

\bibitem{Venkatakrishnan2013}
S.~V. Venkatakrishnan, C.~A. Bouman, and B.~Wohlberg,
\newblock ``Plug-and-play priors for model-based reconstruction,''
\newblock in {\em IEEE Glob. Conf. Signal Inf. Process. Proc.}, Austin, TX,
  USA, 2013, pp. 945--948.

\bibitem{park2025plug}
C.~Y. Park, Y.~Hu, M.~T. McCann, C.~Garcia-Cardona, B.~Wohlberg, and U.~S.
  Kamilov,
\newblock ``Plug-and-play priors as a score-based method,''
\newblock in {\em 2025 IEEE International Conference on Image Processing
  (ICIP)}. IEEE, 2025, pp. 49--54.

\bibitem{hurault2024convergent}
S.~Hurault, A.~Chambolle, A.~Leclaire, and N.~Papadakis,
\newblock ``Convergent plug-and-play with proximal denoiser and unconstrained
  regularization parameter,''
\newblock {\em J. Math. Imaging Vis.}, vol. 66, no. 4, pp. 616--638, 2024.

\bibitem{wang2022}
P.~Wang, B.~Bayram, and E.~Sertel,
\newblock ``A comprehensive review on deep learning based remote sensing image
  super-resolution methods,''
\newblock {\em Earth-Science Reviews}, vol. 232, pp. 104110, 2022.

\bibitem{tao2020}
H.~Tao,
\newblock ``Super-resolution of remote sensing images based on a deep
  plug-and-play framework,''
\newblock in {\em IEEE International Geoscience and Remote Sensing Symposium},
  2020, pp. 625--628.

\bibitem{hurault2022proximal}
S.~Hurault, A.~Leclaire, and N.~Papadakis,
\newblock ``Proximal denoiser for convergent plug-and-play optimization with
  nonconvex regularization,''
\newblock in {\em International Conference on Machine Learning}, 2022, pp.
  9483--9505.

\bibitem{pesquet2021learning}
J.-C. Pesquet, A.~Repetti, M.~Terris, and Y.~Wiaux,
\newblock ``Learning maximally monotone operators for image recovery,''
\newblock {\em SIAM J. Imaging Sci.}, vol. 14, no. 3, pp. 1206--1237, 2021.

\bibitem{combettes2005signal}
P.~L. Combettes and V.~R. Wajs,
\newblock ``Signal recovery by proximal forward-backward splitting,''
\newblock {\em Multiscale Model. Simul.}, vol. 4, no. 4, pp. 1168--1200, 2005.

\bibitem{zhang2021}
K.~Zhang, Y.~Li, W.~Zuo, L.~Zhang, L.~Van Gool, and R.~Timofte,
\newblock ``Plug-and-play image restoration with deep denoiser prior,''
\newblock {\em IEEE Transactions on Pattern Analysis and Machine Intelligence},
  vol. 44, no. 10, pp. 6360--6376, 2021.

\bibitem{ronneberger2015u}
O.~Ronneberger, P.~Fischer, and T.~Brox,
\newblock ``U-net: Convolutional networks for biomedical image segmentation,''
\newblock in {\em International Conference on Medical image computing and
  computer-assisted intervention}. Springer, 2015, pp. 234--241.

\bibitem{he2016deep}
K.~He, X.~Zhang, S.~Ren, and J.~Sun,
\newblock ``Deep residual learning for image recognition,''
\newblock in {\em IEEE/CVF Conference on Computer Vision and Pattern
  Recognition}, 2016, pp. 770--778.

\bibitem{tachella2025deepinverse}
J.~Tachella, M.~Terris, S.~Hurault, A.~Wang, D.~Chen, and et~al.,
\newblock ``Deep{I}nverse: A python package for solving imaging inverse
  problems with deep learning,'' 2025.

\bibitem{pekel2016high}
J.-F. Pekel, A.~Cottam, N.~Gorelick, and A.~S. Belward,
\newblock ``High-resolution mapping of global surface water and its long-term
  changes,''
\newblock {\em Nature}, vol. 540, no. 7633, pp. 418--422, 2016.

\bibitem{efrat2013accurate}
N.~Efrat, D.~Glasner, A.~Apartsin, B.~Nadler, and A.~Levin,
\newblock ``Accurate blur models vs. image priors in single image
  super-resolution,''
\newblock in {\em IEEE International Conference on Computer Vision}, 2013, pp.
  2832--2839.

\bibitem{mallat1999wavelet}
S.~Mallat,
\newblock {\em A wavelet tour of signal processing},
\newblock Elsevier, 1999.

\end{thebibliography}

\end{document}